\documentclass[a4paper,11pt]{article}

\usepackage{apalike}
\usepackage{graphicx}

\renewcommand{\theequation}{\arabic{section}.\arabic{equation}}

\newcommand{\beq}{\begin{equation}}
\newcommand{\eeq}{\end{equation}}
\newcommand{\bea}{\begin{eqnarray}}
\newcommand{\eea}{\end{eqnarray}}

\begin{document}

\title{Does Corticothalamic Feedback Control Cortical Velocity Tuning?}
\author{Ulrich Hillenbrand\thanks{Present address: Institute of Robotics and Mechatronics, German Aerospace Center, Oberpfaffenhofen, 82234 Wessling, Germany. Email: Ulrich.Hillenbrand@dlr.de} and J.\ Leo van Hemmen\\Physik Department, TU M\"unchen\\85747 Garching bei M\"unchen\\Germany}

\date{}

\maketitle

\begin{abstract}
The thalamus is the major gate to the cortex and its contribution to cortical
receptive field properties is well established. Cortical feedback to the
thalamus is, in turn, the anatomically dominant input to relay cells, yet its
influence on thalamic processing has been difficult to interpret. For an
understanding of complex sensory processing, detailed concepts of the
corticothalamic interplay need yet to be established. To study
corticogeniculate processing in a model, we draw on various physiological and
anatomical data concerning the intrinsic dynamics of geniculate relay neurons,
the cortical influence on relay modes, lagged and nonlagged neurons, and the
structure of visual cortical receptive fields. In extensive computer
simulations we elaborate the novel hypothesis that the visual cortex controls
via feedback the temporal response properties of geniculate relay cells in a
way that alters the tuning of cortical cells for speed.
\end{abstract}

\ \\
\noindent Published as {\em Neural Computation} {\bf 13} (2001), pp.\ 327--355.
\ \\

\section{Introduction}

The thalamus is the major gate to the cortex for peripheral sensory signals,
for input from various subcortical sources, and for reentrant cortical
information. Thalamic nuclei, however, do not merely relay information to the
cortex but perform some operation on it while being modulated by various
transmitter systems \cite{McCormick92} and in continuous interplay with their
cortical target areas \cite{Guillery95,Sherman96,Sherman&Guillery96}. Indeed,
cortical feedback to the thalamus is the anatomically dominant input to relay
cells even in those thalamic nuclei that are directly driven by peripheral
sensory systems. While it is well established that the receptive fields of
cortical neurons are strongly influenced by convergent thalamic inputs of
different types
\cite{Saul&Humphrey92a,Saul&Humphrey92b,Reid&Alonso95,Alonso_etal96,Ferster_etal96,Jagadeesh_etal97,Murthy_etal98,Hirsch_etal98},
the modulation effected by cortical feedback in thalamic response has been
difficult to interpret. Experiments and theoretical considerations have pointed
to a variety of operations of the visual cortex on the lateral geniculate
nucleus (LGN), such as attention-related gating of geniculate relay cells
(GRCs) \cite{Sherman&Koch86}, gain control of GRCs \cite{Koch87}, synchronizing
firing of neighboring GRCs \cite{Sillito_etal94,Singer94}, increasing mutual
information between GRCs' retinal input and their output
\cite{McClurkin_etal94}, and switching GRCs from a detection to an analyzing
mode \cite{Godwin_etal96,Sherman96,Sherman&Guillery96}. Nonetheless, the
evidence for any particular function to date is still sparse and rather
indirect.

Clearly, detailed concepts of the interdependency of thalamic and cortical
operation could greatly advance our ideas about complex sensory, and ultimately
cognitive, processing. Here we present a novel view on the corticothalamic
puzzle by proposing that control of velocity tuning of visual cortical
neurons may be an eminent function of corticogeniculate processing. We have outlined some of the ideas in Hillenbrand \& van Hemmen (2000)\nocite{Hillenbrand&vanHemmen00} previously.

In this section we will review facts, some well established, others still
controversial, on the thalamocortical system in order to clear the ground for
the following simulation of the primary visual pathway.

\subsection{Geniculate Response Timing and Cortical Velocity\\Tuning}
\label{I.1}

Velocity selectivity or velocity tuning, taken here to mean preference for a
certain speed and direction of motion of visual features, requires convergence
of pathways with different spatial information and different temporal
characteristics, such as delays, onto single neurons; see, e.g., Hassenstein \&
Reichardt (1956), Watson \& Ahumada (1985), Emerson
(1997)\nocite{Hassenstein&Reichardt56,Watson&Ahumada85,Emerson97}. For higher
mammals this is believed to occur in the primary visual cortex
\cite{Movshon75,Orban_etal81a,Orban_etal81b}.

\begin{figure}
\centerline{\includegraphics[width=\textwidth]{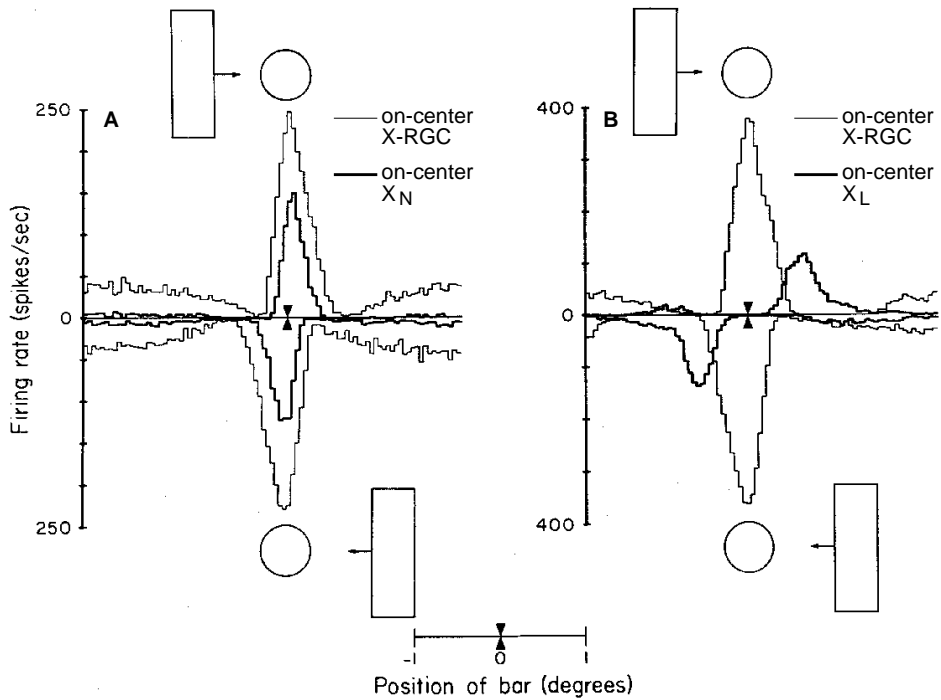}}
\caption[]{Averaged responses of nonlagged ({\bf A}, X$_{\rm N}$, thick line)
and lagged ({\bf B}, X$_{\rm L}$, thick line) geniculate on-center X-cells and
their respective main excitatory retinal input ({\bf A} and {\bf B}, X-RGC,
thin lines) to a moving light bar. Upper and lower histograms show responses to
opposite directions of motion. Double arrowheads indicate the position of the
central point of the receptive fields, circles indicate the approximate size of
the receptive-field centers. The width of the bar was 0.5 degrees and is drawn
to scale. The bar was swept at 5 deg$/$s and 100 times for the X$_{\rm N}$-cell,
102 times for the X$_{\rm L}$-cell in each direction. Spikes were collected in
bins of 10 ms width. Figure adapted from Mastronarde
(1987a)\nocite{Mastronarde87a}.}
\label{Fig_Mastronarde_87_bar}
\end{figure}

In the A-laminae of cat LGN two types of X-relay cell have been identified that
dramatically differ in their temporal response properties
\cite{Mastronarde87a,Humphrey&Weller88a,Saul&Humphrey90}. Those that are more
delayed in response time and phase have been termed {\em lagged}, the others
{\em nonlagged} cells (with the exception of very few so-called {\em partially
lagged} neurons); see Figure \ref{Fig_Mastronarde_87_bar}. In lagged neurons,
the on-response to a flash of light is preceded by a dip in the firing rate
lasting for 5 to 220 ms and there is typically a transient of high firing rate
just after the offset of a prolonged light stimulus
\cite{Mastronarde87a,Humphrey&Weller88a}. For a moving light bar, the time lag
of the on-response peak is about 100 ms after the bar has passed the receptive
field (RF) center \cite{Mastronarde87a}. In contrast, the nonlagged cells'
responses resemble their retinal input and show no transient at stimulus offset
\cite{Mastronarde87a,Humphrey&Weller88a}. Lagged X-cells comprise about 40 \%
of all X-relay cells \cite{Mastronarde87a,Humphrey&Weller88b}. Physiological
\cite{Mastronarde87b}, pharmacological \cite{Heggelund&Hartveit90}, and
structural \cite{Humphrey&Weller88b} evidence suggests that rapid feedforward
inhibition via intrageniculate interneurons plays a decisive role in shaping
the lagged cells' response. Some authors have additionally related differences
in receptor types to the lagged-nonlagged dichotomy
\cite{Heggelund&Hartveit90,Hartveit&Heggelund90}; see, however, Kwon et al.\
(1991)\nocite{Kwon_etal91}.

Layer 4B in cortical area 17 of the cat is the target of both lagged and
nonlagged geniculate X-cells
\cite{Saul&Humphrey92a,Jagadeesh_etal97,Murthy_etal98}. The {\em
spatiotemporal} RFs of its direction-selective simple cells can routinely be
interpreted as being composed of subregions that receive geniculate inputs
alternating between lagged and nonlagged X-type
\cite{Saul&Humphrey92a,Saul&Humphrey92b,DeAngelis_etal95,Jagadeesh_etal97,Murthy_etal98},
just as convergent and segregated geniculate on- and off-inputs have been shown
to outline the {\em spatial} structure of the simple cells' RFs
\cite{Reid&Alonso95,Alonso_etal96,Ferster_etal96,Hirsch_etal98}. According to
this view, directional selectivity is created by the response-phase difference
of roughly a {\em quarter cycle} between successive off-lagged, off-nonlagged,
on-lagged, and on-nonlagged responses across the RF. At least for simple cells
in layer 4B, this RF structure determines the response to moving visual
features
\cite{McLean&Palmer89,McLean_etal94,Reid_etal91,Albrecht&Geisler91,DeAngelis_etal93,DeAngelis_etal95,Jagadeesh_etal93,Jagadeesh_etal97,Murthy_etal98},
and hence the cell's tuning for direction and speed\footnote{To avoid
confusion, we point out that the term `speed tuning' is sometimes used in a
more restricted sense. Simple cells exhibit tuning for spatial and temporal
frequencies that results in preference for speeds of moving gratings depending
on their spatial frequency. Here we will be concerned with the more natural
case of stimuli having a low-pass frequency content \cite{Field94},
specifically, those composed of local features such as thin bars.}. Lagged and
nonlagged inputs that converge, either directly or via other cortical neurons,
on simple cells and segregate in separate subregions of the simple cells' RF
are thus likely to contribute to the earliest level of cortical velocity
selectivity
\cite{Saul&Humphrey90,Saul&Humphrey92a,Saul&Humphrey92b,Ferster_etal96,Jagadeesh_etal97,Wimbauer_etal97a,Wimbauer_etal97b,Murthy_etal98}.

Certainly, intracortical input to cortical cells also contributes to
velocity-selective responses, given that these inputs anatomically outnumber
thalamic inputs \cite{Ahmed_etal94}. Suggested intracortical effects include
sharpening of tuning properties by suppressive interactions
\cite{Hammond&Pomfrett90,Reid_etal91,Hirsch_etal98,Crook_etal98,Murthy&Humphrey99},
amplification of geniculate inputs by recurrent excitation
\cite{Douglas_etal95,Suarez_etal95}, and normalization of responses by local
interactions \cite{Toth_etal97}. Intracortical circuits can in principle even
generate their own direction selectivity by selectively inhibiting responses to
nonpreferred motion \cite{Douglas_etal95,Suarez_etal95,Maex&Orban96}. Our
modeling is complementary to the latter in that we emphasize the influence of
geniculate inputs on cortical RF properties that is suggested by numerous
studies
\cite{Saul&Humphrey92a,Saul&Humphrey92b,Reid&Alonso95,Alonso_etal96,Ferster_etal96,Toth_etal97,Jagadeesh_etal97,Murthy_etal98,Hirsch_etal98},
in order to bring out effects that are specific to the geniculate contribution
to spatiotemporal tuning.

Great care must be taken when extrapolating from cats to primates. In
particular, no lagged relay cells have been described in the primate LGN so
far. On the other hand, a recent study \cite{DeValois&Cottaris98} does suggest
a set of geniculate inputs to directionally selective simple cells in macaque
striate cortex that is essentially analogous, in terms of response properties,
to the lagged-nonlagged set envisaged for cat simple cells. If the underlying
physiology for primates turns out to be similar to the one for cats, the
results presented here extend to primates.

\subsection{Geniculate Relay Modes}
\label{I.2}

Thalamocortical neurons possess a characteristic blend of voltage-gated ion
channels
\cite{Jahnsen&Llinas84a,Jahnsen&Llinas84b,Huguenard&McCormick92,McCormick&Huguenard92}
that jointly determine the timing and pattern of action potentials in response
to a sensory stimulus; see the Appendix for a brief introduction to models of
ion currents. Depending on the initial membrane polarization, the GRC response
to a visual stimulus is in a range between a {\em tonic} and a {\em burst} mode
\cite{Sherman96,Sherman&Guillery96}. At hyperpolarization below roughly -70 mV,
a Ca$^{2+}$ current, called the low-threshold Ca$^{2+}$ current or T-current
($I_{\rm T}$; T for `transient'), gets slowly de-inactivated. As the membrane
depolarizes above roughly -70 mV, the current activates, followed by a rapid
transition from the de-inactivated to the inactivated state, thereby producing
a Ca$^{2+}$ spike with an amplitude that depends on how long and how strongly
the cell has been hyperpolarized previously. After sufficient hyperpolarization
the Ca$^{2+}$ spike will thus reach the threshold for Na$^+$ spiking and give
rise to a burst of one to seven action potentials riding its crest
\cite{Jahnsen&Llinas84a,Jahnsen&Llinas84b,Huguenard&McCormick92,McCormick&Huguenard92}.
All other action potentials, i.e., those that are not promoted by a Ca$^{2+}$
spike and, hence, do not group into bursts, are called tonic spikes.

Although the issue is still controversial, there is some evidence that a {\em
mixture} of burst and tonic spikes may be involved in the transmission of
visual signals in lightly anesthetized or awake animals
\cite{Guido_etal92,Guido_etal95,Guido&Weyand95,Mukherjee&Kaplan95,Sherman96,Sherman&Guillery96,Reinagel_etal99}.
In lagged cells, because of the strong feedforward inhibition they are assumed
to receive, burst spikes have been held responsible for the high-activity
transient seen at the offset of their retinal input
\cite{Mastronarde87a,Mastronarde87b}, thus contributing substantially to the
delayed peak response to a moving bar \cite{Mastronarde87b}. In nonlagged cells
at resting membrane potentials below -70 mV, bursting constitutes a very early
part of the visual response, producing a phase lead of up to a quarter cycle
relative to their retinal input
\cite{Lu_etal92,Guido_etal92,Mukherjee&Kaplan95}.  At more depolarized membrane
potentials nonlagged responses are dominated by tonic spikes and are in phase
with retinal input \cite{Lu_etal92,Guido_etal92,Mukherjee&Kaplan95}.

Cortical feedback to the A-laminae of the LGN, arising mainly from layer 6 of
area 17 \cite{Sherman96,Sherman&Guillery96}, can locally modulate the response
mode, and hence the {\em timing}, of GRCs by shifting their membrane potentials
on a time scale that is long as compared to retinal inputs.  This may occur
directly through the action of metabotropic glutamate and NMDA receptors
(depolarization)
\cite{McCormick&vonKrosigk92,Godwin_etal96,Sherman96,Sherman&Guillery96,vonKrosigk_etal99}
and indirectly via the perigeniculate nucleus (PGN) or geniculate interneurons
by activation of GABA$_{\rm B}$ receptors (hyperpolarization) of GRCs
\cite{Crunelli&Leresche91,Sherman&Guillery96,vonKrosigk_etal99}. Indeed, GRCs
in vivo are dynamic and differ individually in their degree of burstiness
\cite{Lu_etal92,Guido_etal92,Mukherjee&Kaplan95}. Here we explicate the causal
link between the {\em variable response timing of GRCs} and {\em variable
tuning of cortical simple cells for speed} of moving features, thus identifying
control of speed tuning as a likely mode of corticothalamic operation.

\section{The Model}

Before turning to our simulation results we describe in this section the
underlying model of the primary visual pathway.

\subsection{Geniculate Input to the Primary Visual Cortex}
\label{Gen_Input}

\begin{figure}
\begin{minipage}{0.47\textwidth}
\leftline{\includegraphics[width=0.9\textwidth]{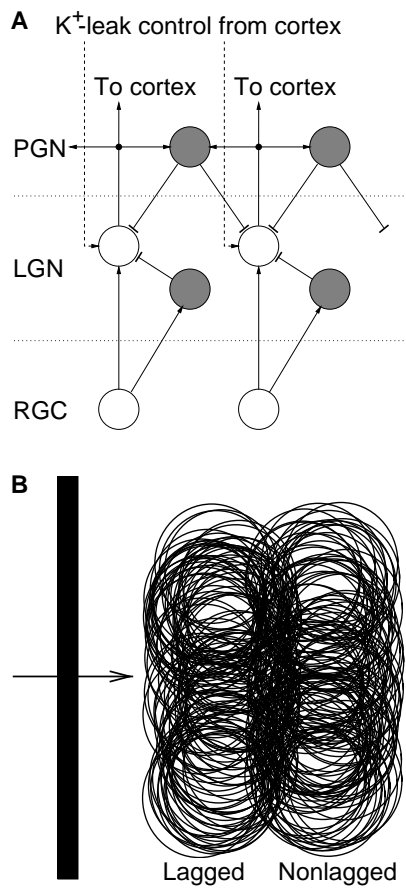}}
\end{minipage}
\begin{minipage}{0.52\textwidth}
\caption[]{Model of the primary visual pathway. ({\bf A}) Open/filled circles
and arrow/bar heads indicate excitatory/inhibitory neurons and their respective
synapses. A retinal ganglion cell (RGC) sends its axon to the lateral
geniculate nucleus (LGN) and synapses excitatorily on a relay cell (open
circle) and on an intrageniculate interneuron (filled circle), which in turn
inhibits the same relay cell (arrangement called `synaptic triad'). The
relative strengths of feedforward excitation and feedforward inhibition shape a
relay cell's response to be of the lagged or nonlagged type (see main text and
Figure \ref{lnl_synphases}). There is an inhibitory feedback loop via the
perigeniculate nucleus (PGN). The influence of cortical feedback has been
modeled as a variation of the relay cells' resting membrane potential by
control of a K$^+$ leak current. There is no cortical input to the PGN in the
model. Moreover, we neglect any fast (ionotropic) cortical feedback. Possible
effects of such feedback are discussed in section \ref{Discussion}. ({\bf B})
Arrangement in visual space of the receptive field (RF) centers of the 100
lagged and 100 nonlagged relay cells comprising the model LGN. These relay
cells are envisaged to project onto the same cortical simple cell and create an
on- or off-region of its RF. In the simulations, the diameter of a single
lagged or nonlagged RF center is 0.5 degrees. Results for rescaled versions of
this geometry can be derived straightforwardly from the simulations; see
section \ref{Discussion}. The bar and arrow on the left indicate
preferred orientation and direction of motion, respectively. Figure adapted
from Hillenbrand \& van Hemmen (2000)\nocite{Hillenbrand&vanHemmen00}.}
\label{model_graphs}
\end{minipage}
\end{figure}

For the GRCs we have employed a 12-channel model of the cat relay neuron
\cite{Huguenard&McCormick92,McCormick&Huguenard92}, adapted to 37 degrees
Celsius; see the Appendix for a brief introduction to biophysical neuron
models. The neuron model includes a transient and a persistent Na$^+$ current,
several voltage-gated K$^+$ currents, a voltage- and Ca$^{2+}$-gated K$^+$
current, a low- and a high-threshold Ca$^{2+}$ current, a
hyperpolarization-activated mixed cation current, and Na$^+$ and K$^+$ leak
conductances. As shown in Figure \ref{model_graphs}{\bf A}, retinal input
reaches a GRC directly as excitation, and indirectly via an intrageniculate
interneuron as inhibition, thus establishing the typical {\em triadic} synaptic
circuit found in the glomeruli of X-GRCs
\cite{Sherman&Koch90,Sherman&Guillery96}. The temporal difference between the
two afferent pathways equals the delay of the inhibitory synapse and has been
taken to be 1.0 ms \cite{Mastronarde87b}.

As will be described in more detail below, we have found typical lagged
responses for strong feedforward inhibition with weak feedforward excitation,
in agreement with Mastronarde (1987b), Humphrey \& Weller (1988b), and
Heggelund \& Hartveit
(1990)\nocite{Mastronarde87b,Humphrey&Weller88b,Heggelund&Hartveit90}. On the
other hand, typical nonlagged responses are produced by weak feedforward 
inhibition with strong feedforward excitation. We have therefore implemented
lagged and nonlagged relay cells in the model by varying the relative strengths
of feedforward excitation and feedforward inhibition.

It is known that both NMDA and non-NMDA receptors contribute to
retinogeniculate excitation to varying degrees, ranging from almost pure
non-NMDA to almost pure NMDA-mediated responses in individual GRCs of both
lagged and nonlagged varieties \cite{Kwon_etal91}. At least in lagged cells,
however, early responses and, hence, responses to the transient stimuli that
will be considered here, seem to depend to a lesser degree on the NMDA receptor
type than late responses \cite{Kwon_etal91}. Since the essential
characteristics of lagged and nonlagged responses apparently do not depend on
the special properties of NMDA receptors -- an assumption confirmed by our
results -- we have chosen the postsynaptic conductances in GRCs to be entirely
of the non-NMDA type.

The time course of postsynaptic conductance change in GRCs following reception
of an input has been modeled by an alpha function,
\beq
{\rm g}(t>0) \, = \, g_{\rm max} \, \frac{t}{\tau} \, \exp\!\left(1 - \frac{t}{\tau}\right) \ .
\label{g(t)}
\eeq
For excitation, the rise time $\tau$ has been chosen to be 0.4 ms
\cite{Mukherjee&Kaplan95}, for inhibition it is 0.8 ms. The latter value was
estimated from the relative durations of S potentials recorded at excitatory
and inhibitory geniculate synapses \cite{Mastronarde87b} and was found to
reproduce the rise times of inhibitory postsynaptic potentials recorded in
relay cells following stimulation of the optic chiasm
\cite{Bloomfield&Sherman88}. The reversal potentials are for excitation 0 mV,
for inhibition $-$85.8 mV \cite{Bal_etal95a}.

The model system comprises 100 lagged and 100 nonlagged relay neurons. Their RF
centers are 0.5 degrees in diameter \cite{Cleland_etal79} and are spatially
arranged in a lagged and a nonlagged cluster subtending 0.7 degrees each and
displaced by 0.45 degrees; see Figure \ref{model_graphs}{\bf B}. More
precisely, the central points of the RFs of lagged and nonlagged cells are
uniformly distributed within two separate intervals of 0.2 degrees each along a
certain axis, which will be the axis of bar motion during stimulation; see
section \ref{Stimulation}. The RFs' offsets in the direction orthogonal to this
axis, i.e., in the direction that defines the preferred orientation of the bulk
RF, are irrelevant as long as the stimulus bar is long enough to pass through
all RFs of the relay cells in one sweep. In fact, the bar used in the
simulations is much longer than typical RFs of simple cells; see section
\ref{Stimulation}.

The layout of geniculate inputs (Figure \ref{model_graphs}{\bf B}) matches the
basic structure of a single on- or off-region in an RF of a directional simple
cell in cortical layer 4B onto which the GRCs are envisaged to project
\cite{Saul&Humphrey92a,Saul&Humphrey92b,DeAngelis_etal95,Jagadeesh_etal97,Murthy_etal98}.
To complete the geniculate input to a RF of this type, this lagged-nonlagged
unit would have to be repeated with alternating on-off-polarity and a spatial
offset that would determine the simple cell's preference for some spatial
frequency. Since we are not concerned here with effects of spatial frequency
(see previous footnote 1), omission of the other on/off-regions does not affect
our conclusions. Results for rescaled RF geometries can be derived
straightforwardly from the simulations; see section \ref{Discussion}.

The number of geniculate cells contributing to a simple cell's RF has been
estimated roughly from Ahmed et al.\ (1994)\nocite{Ahmed_etal94}. Only its
order of magnitude matters.

We have also taken into account feedback inhibition via the PGN
\cite{Lo&Sherman94,Sherman&Guillery96}; see Figure \ref{model_graphs}{\bf A}.
Connections between PGN neurons and GRCs are all to all within, and separate
for the lagged and nonlagged populations\footnote{This synaptic separation of
the lagged and the nonlagged pathways was implemented solely to allow for
independent simulation of the two. Although an inhibitory coupling of lagged
and nonlagged cells could in reality cause some anti-correlation of their
firing, there is no evidence for anti-correlation of GRCs. Any such effects
thus seem negligible. In any case, they would not affect our conclusions.}.
Axonal plus synaptic delays are 2.0 ms in both directions.

Intrageniculate interneurons and PGN cells, like GRCs, possess a complex blend
of ionic currents. They are, however, thought to be active mainly in a tonic
spiking mode during the awake state \cite{Contreras_etal93,Pape_etal94}. For an
efficient usage of computational resources and time we have therefore modeled
these neurons by the spike-response model \cite{Gerstner&vanHemmen92}, which
gives a reasonable approximation to tonic spiking \cite{Kistler_etal97}. Note
that for the present model it is irrelevant whether transmission across
dendrodendritic synapses between intrageniculate interneurons and GRCs actually
occurs with or without spikes; cf.\ Cox et al.\ (1998)\nocite{Cox_etal98}. For
a relay neuron, all that matters is the fact that an excitatory retinal input
is mostly followed by an inhibitory input \cite{Bloomfield&Sherman88}. The
spike-response neurons have been given an adaptive spike output, implemented as
an accumulating refractory potential \cite{Gerstner&vanHemmen92}, i.e., there
is some adaptation of transmission across the dendrodendritic synapses. The
refractory potential, and hence the effect of adaptation, saturates on a time
scale of 10.0 ms.

\subsection{Cortical Feedback}
\label{CTX_feedback}

Metabotropic glutamate receptors effect a closing of K$^+$ leak channels and a
membrane depolarization, while GABA$_{\rm B}$ receptors, via PGN or geniculate
interneurons, effect an opening of K$^+$ leak channels and a membrane
hyperpolarization. Accordingly, we have incorporated the influence of cortical
feedback to the thalamus by varying the K$^+$ leak conductance of GRCs
\cite{McCormick&vonKrosigk92,Godwin_etal96}; see Figure \ref{model_graphs}{\bf
A}. The resulting stationary membrane potential in the absence of any retinal
input will be called {\em resting membrane potential}. All GRCs, lagged and
nonlagged, have been assigned the same resting membrane potential; here we
assume a uniform action of cortical feedback at least on the scale of single
RFs in area 17. By varying the resting membrane potential we investigate a
strictly modulatory role of corticogeniculate feedback, as opposed to the
retinal inputs that drive relay cells to fire; cf.\ Sherman \& Guillery (1996),
Crick \& Koch (1998)\nocite{Sherman&Guillery96,Crick&Koch98}.

For every single stimulus presentation (see below) we have kept the K$^+$ leak
conductance constant. This is justified by the slow action, compared to typical
passage times of local stimulus features through RFs, of the metabotropic
receptors, ranging from hundreds of milliseconds for GABA$_{\rm B}$ to seconds
for metabotropic glutamate receptors \cite{vonKrosigk_etal99}. Nonetheless, it
is clear that dynamics in the corticogeniculate pathway may produce effects for
slow-moving stimuli that we here cannot account for.

In modeling cortical feedback we neglect input to the LGN that is mediated by
ionotropic receptors and, hence, acts on a much shorter timescale. Furthermore,
we do not explicitly model cortical input to the PGN. The effects those inputs
may have on our results are discussed in section \ref{Discussion}.

\subsection{Stimulation}
\label{Stimulation}

The input to GRCs has been modeled as a set of Poisson spike trains with
time-varying firing rates. For investigation of the temporal transfer
characteristics of lagged and nonlagged neurons, these rates varied sinusoidally
between 0 and 100 spikes$/$s (amplitude 50 spikes$/$s, DC component 50
spikes$/$s) at a range of temporal frequencies. Before any responses to
sinusoidal stimuli have been collected, the stimuli were presented for 1
second, that is, depending on the frequency, between 1 and 11 cycles. We have
recorded the response for the following 100 seconds of stimulus presentation.

\begin{figure}
\begin{minipage}{0.45\textwidth}
\leftline{\includegraphics[width=0.9\textwidth]{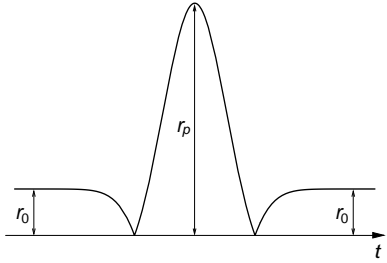}}
\end{minipage}
\begin{minipage}{0.54\textwidth}
\caption[]{Time-dependent rate response to a moving bar of a retinal ganglion
cell; cf.\ equation \ref{bar_rate}. This firing rate has been used in the
simulations to generate input spikes to geniculate relay cells by an
inhomogeneous Poisson process. The peak rate $r_p$ and the width of the
response peak have been adjusted for different bar speeds to fit the data of
Cleland \& Harding (1983)\nocite{Cleland&Harding83}. The background rate $r_0$
is taken to be 38 spikes$/$s \cite{Mastronarde87b}.}
\label{RGC_bar_resp_fit}
\end{minipage}
\end{figure}

For studying the responses to moving bars, rates have been fitted to recordings
from retinal ganglion cells in response to moving, thin (0.1 degrees), long (10
degrees) bars \cite{Cleland&Harding83}. The fit for a single retinal ganglion
cell is of the form
\beq
{\rm r}(t) \, = \, \left| (r_p + r_0) \,
\exp\!\left[-\left(\frac{t}{\Delta}\right)^2\right] - r_0 \right| \ ,
\label{bar_rate}
\eeq
where $r_p$ is the peak rate, $r_0$ is the background rate, and $\Delta$ is a
width parameter; see Figure \ref{RGC_bar_resp_fit}. For the different speeds of
bar motion used in the simulations, $r_p$ and $\Delta$ have been chosen to fit
the data of Cleland \& Harding (1983)\nocite{Cleland&Harding83} while $r_0 =
38$ spikes$/$s \cite{Mastronarde87b}. Different GRCs received retinal input from
statistically independent sources.

We have studied bar responses of single lagged and nonlagged neurons as well as
of the entire population of 100 lagged and 100 nonlagged neurons in the
geniculate model. Accordingly, bars were moved across single RFs of relay cells
or the whole bulk RF in the preferred and anti-preferred directions; see Figure
\ref{model_graphs}{\bf B}. Bar motion always started $3 \, \Delta$ (cf.\
equation \ref{bar_rate}) before it hit the first RF center, and stopped $3 \,
\Delta$ after it had passed the last RF center. There was a 1 second interval
of stimulation with the background activity (38 spikes$/$s) between bar sweeps.

\subsection{Data Analysis}
\label{DataAnalysis}

We collected spike times with 0.1 ms resolution. Spikes of single relay neurons
in response to moving bars were counted in bins of 5 ms, a timescale relevant
to postsynaptic integration, for variable synaptic excitatory and inhibitory
input strengths. The bin counts have been averaged over 100 bar sweeps at each
velocity and synaptic setting.

Spikes of single lagged and nonlagged neurons in response to sinusoidal
stimulation with variable frequency $\omega$ were counted in a time window of 5
ms shifted by steps of 1 ms. The spike counts were averaged over all cycles of
the stimulus presented within 100 seconds of stimulation. For the resulting
spike-count functions we determined the amplitude $F_1(\omega)$ and the phase
$\phi_1(\omega)$ of their first Fourier component. With the amplitude $A$ ($=
50$ spikes$/$s; see section \ref{Stimulation}) and the phase $\psi$ of the
sinusoidal input rate, we have calculated the amplitude-transfer function
$F_1(\omega)/A$ and the phase-transfer function $\psi - \phi_1(\omega)$;
negative phase transfer means phase lead over the input.

For the investigation of velocity tuning, spikes of all 100 lagged and 100
nonlagged relay cells were pooled. For each velocity $v$ of bar motion tested,
we calculated the total lagged and nonlagged response rates ${\rm r}_{\rm
l}(v,t)$ and ${\rm r}_{\rm nl}(v,t)$, respectively, as spike counts in 5 ms
windows shifted by steps of 1 ms ($t=1,2,\ldots$ ms). The velocity tuning of
the pooled lagged and nonlagged peak rates per neuron is
\begin{equation}
{\rm R}_{\ell}(v) \, = \, \frac{1}{100} \, \max_{t \in [t_{\rm i},t_{\rm f}]}
{\rm r}_{\ell}(v,t) \ ,
\quad \ell = {\rm l},{\rm nl} \ ,
\end{equation}
where the times $t_{\rm i}$ and $t_{\rm f}$ are chosen such that all of the
response to a bar sweep lies in the interval $[t_{\rm i},t_{\rm f}]$.

We are primarily interested in the {\em total geniculate input} to a cortical
simple cell onto which the GRCs are envisaged to project. To this end, we
shifted lagged spikes by 2 ms to later times in order to account for the fact
that the lagged cells' conduction times to cortex are slightly longer than
those of the nonlagged cells \cite{Mastronarde87a,Humphrey&Weller88a}.
Furthermore, although lagged responses in the LGN tend to be weaker than
nonlagged responses \cite{Mastronarde87a,Humphrey&Weller88a,Saul&Humphrey90},
they appear to be about equally efficient in driving cortical simple cells
\cite{Saul&Humphrey92a}. The cortical (possibly synaptic) cause being beyond
the scope of this work, we simply counted every lagged spike twice to
obtain the velocity tuning of the effective geniculate input to a cortical
cell,
\begin{equation}
{\rm R}(v) \, = \, \frac{1}{100} \, \max_{t \in [t_{\rm i},t_{\rm f}]} \left[
2 \, {\rm r}_{\rm l}(v,t - 2 \, {\rm ms}) + {\rm r}_{\rm nl}(v,t) \right] \ .
\end{equation}
The peak input rate ${\rm R}(v)$ per lagged-nonlagged pair is correlated with
simple-cell activity because postsynaptic potentials are summed almost linearly
in simple cells \cite{Jagadeesh_etal93,Kontsevich95,Jagadeesh_etal97}.

The total geniculate input rate ${\rm R}(v)$ to a cortical neuron depends on
(i) the magnitude of the pooled lagged and nonlagged response peaks, ${\rm
R}_{\rm l}(v)$ and ${\rm R}_{\rm nl}(v)$, respectively, and (ii) their relative
timing. To differentiate between these two factors we determined the times
${\rm t}_{\rm l}(v)$ and ${\rm t}_{\rm nl}(v)$ of the maxima of the lagged and
nonlagged response rates, respectively,
\beq
{\rm t}_{\ell}(v) \, = \, {\rm arg} \max_{t \in [t_{\rm i},t_{\rm f}]}
{\rm r}_{\ell}(v,t) \ ,
\quad \ell = {\rm l},{\rm nl} \ ,
\eeq
and calculated the peak-time differences ${\rm t}_{\rm nl}(v) - {\rm
t}_{\rm l}(v)$ as a function of the bar velocity $v$. Means and standard errors
have been estimated from a sample of 30 bar sweeps at each bar velocity.

\subsection{Numerics}

The model is described by a high-dimensional system of nonlinear, coupled,
stochastic differential equations. For numerical integration of the GRC
dynamics we used an adaptive fifth-order Runge-Kutta algorithm\footnote{An
algorithm for numeric integration of ordinary differential equations is said to
be of $n$th order, if the error per time step $\delta t$ is of order $\delta
t^{n+1}$. Note that, because of discontinuities in the system of differential
equations \cite{Huguenard&McCormick92,McCormick&Huguenard92}, more
sophisticated and faster methods for integration than Runge-Kutta cannot be
safely applied.} \cite{Press_etal92}. The maximal time step was 0.02 ms and was
scaled down to satisfy upper bounds on the estimated error per time step.
Increasing or decreasing those bounds by a factor of 10 had negligible effects
on the time course of the membrane potential of a GRC, and no effect on spike
timing within the temporal resolution of 0.1 ms we used for recording. The
dynamics of spike-response neurons was solved by exact integrals.

Each simulation started with a 3 second period without any stimulus to allow
the GRCs' dynamics to converge on its stationary (resting) state. Simulations
were run on an IBM SP2 parallel computer.

\section{Results}
\label{Results}

We first address the response properties of single relay neurons in the model,
and then turn to the total geniculate input to a cortical neuron.

\subsection{Lagged and Nonlagged Relay Neurons}

\begin{figure}
\centerline{\includegraphics[width=0.97\textwidth]{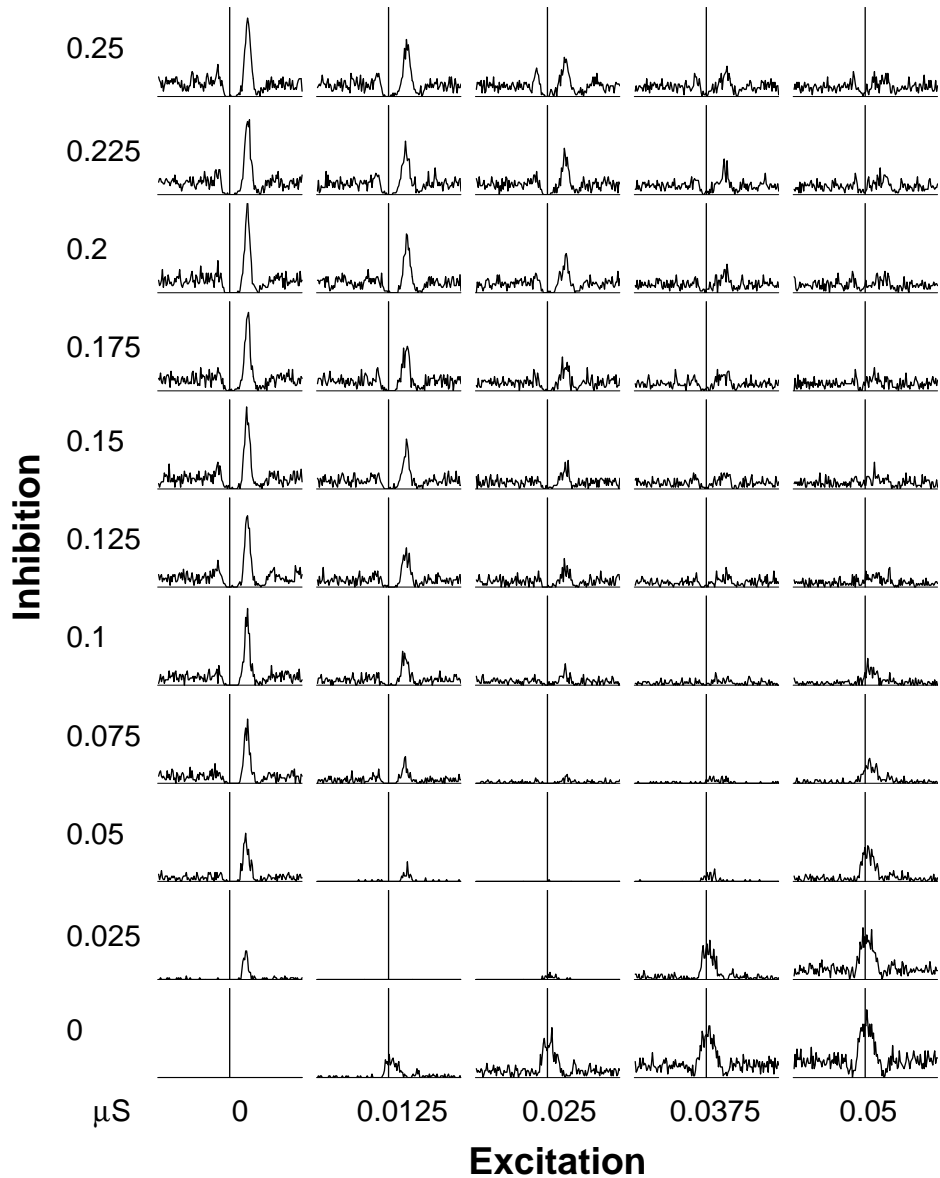}}
\caption[]{Dependence of moving-bar response of single modeled relay neurons
on the strengths of feedforward excitation and feedforward inhibition. In
each plot the horizontal axis spans 750 ms; the vertical axis indicates the
time of the retinal input peak and spans 100 spikes$/$s. Across the whole array
of plots the peak postsynaptic conductances $g_{\rm max}$ (equation \ref{g(t)})
vary for excitation horizontally from 0 to 0.05 $\mu$S, and for inhibition
vertically from 0 to 0.25 $\mu$S. The regions of lagged- and nonlagged-type
responses in this parameter space are at low excitation with high inhibition
and at high excitation with low inhibition, respectively. The resting membrane
potential is $-$65 mV. Responses are averaged over 100 bar sweeps.}
\label{lnl_synphases}
\end{figure}

We have checked whether both lagged- and nonlagged-type responses could be
produced within our model by simply varying the synaptic strengths of
feedforward excitation and feedforward inhibition of relay neurons; see Figure
\ref{model_graphs}{\bf A}. Varying the peak postsynaptic conductances $g_{\rm
max}$ (equation \ref{g(t)}) for excitation and inhibition and stimulating with
a bar moving at 4 deg$/$s we found a lagged-nonlagged transition that is
analogous to a first-order phase transition in response timing; see Figure
\ref{lnl_synphases} for an example at a resting membrane potential of $-$65 mV.
At strong excitation and weak inhibition there is a response peak with zero
delay relative to the input peak. As the excitation is reduced and the
inhibition increased, this nonlagged peak shrinks while a lagged peak develops.
The latter invariably has a delay of roughly 100 ms relative to the input peak,
a value consistent with experimental data \cite{Mastronarde87a}; cf.\ Figure
\ref{Fig_Mastronarde_87_bar}{\bf B}. At strong inhibition and weak excitation
the lagged peak is the dominant part of the response.

\begin{figure}
\centerline{\includegraphics[width=\textwidth]{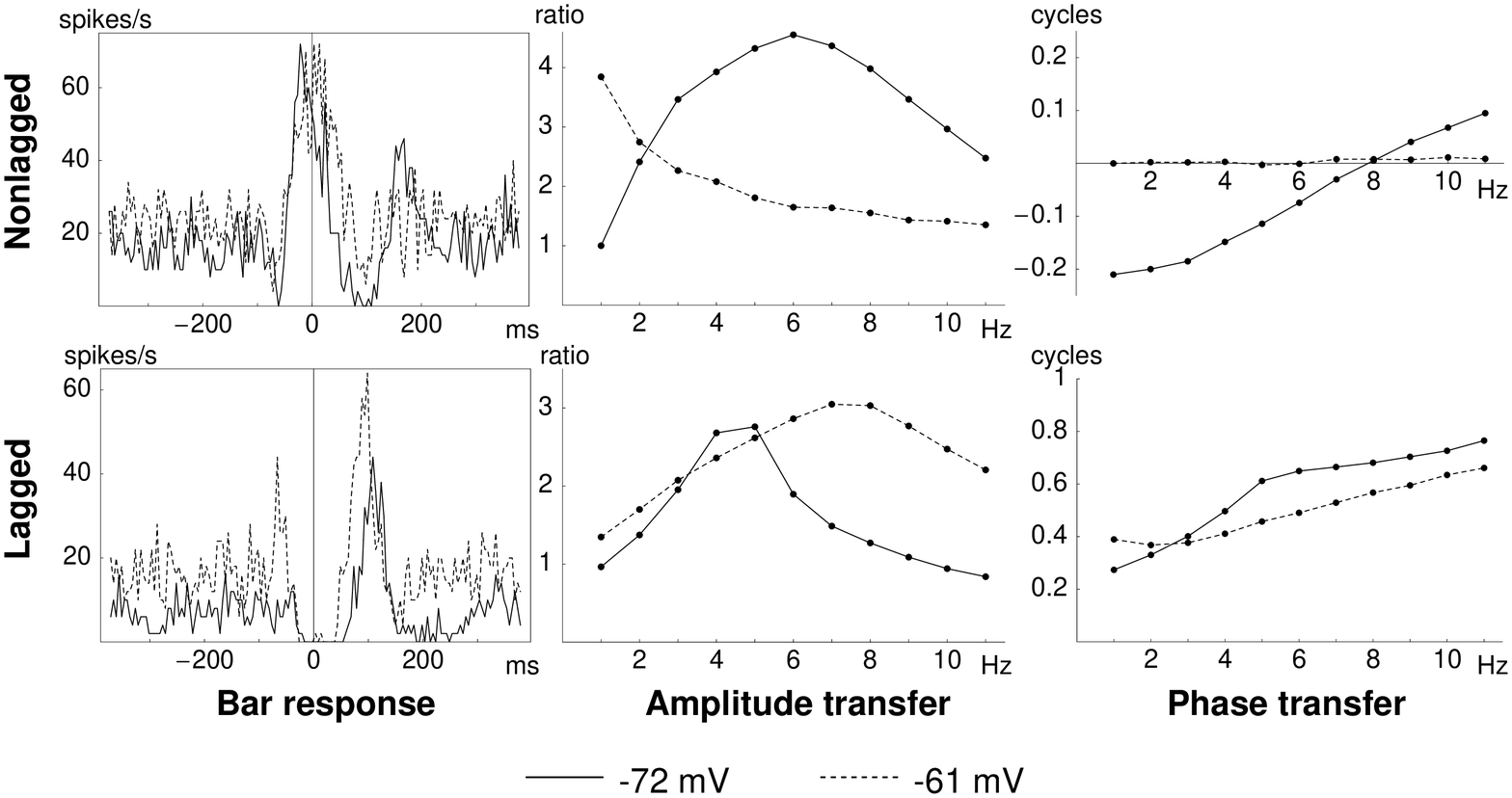}}
\caption[]{Dependence of moving-bar response and temporal transfer function of
single modeled relay neurons on their resting membrane potential. Typical
nonlagged responses (top row, $g_{\rm max} = 0.05$ $\mu$S for excitation and
0.0125 $\mu$S for inhibition) and lagged responses (bottom row, $g_{\rm max} =
0.0125$ $\mu$S for excitation and 0.25 $\mu$S for inhibition; cf.\ Figure
\ref{lnl_synphases}) have been reproduced at the two resting membrane
potentials $-$72 mV (solid lines) and $-$61 mV (dashed lines). For the bar
responses (leftmost column) the time of the retinal input peak has been set to
zero. As the membrane is {\em hyperpolarized}, the nonlagged bar-response peak
shifts to {\em earlier} times. Conversely, the lagged bar response shifts to
{\em later} times. The changes in bar-response timing are also reflected in
corresponding changes in the phase-transfer functions (rightmost column). Bar
responses are averaged over 100 bar sweeps. Amplitude and phase transfer have
been calculated from responses to sinusoidal input rates, averaged over 100
seconds. Note the different scales on the ``cycles'' axes for nonlagged and
lagged cells.}
\label{61_72_mV}
\end{figure}

We have also checked the dependence of relay-cell responses on their resting
membrane potential. The peak postsynaptic conductances $g_{\rm max}$ for the
lagged cell have now been fixed at 0.0125 $\mu$S for excitation and at 0.25
$\mu$S for inhibition; for the nonlagged cell they have been fixed at 0.05
$\mu$S for excitation and at 0.0125 $\mu$S for inhibition; cf.\ Figure
\ref{lnl_synphases}. In Figure \ref{61_72_mV} we show the bar response (4
deg$/$s) and the temporal transfer of amplitude and phase of a lagged and a
nonlagged neuron for the resting membrane potentials $-$72 mV and $-$61 mV. 
Again, the response data agree well with experiments
\cite{Mastronarde87a,Saul&Humphrey90,Lu_etal92,Guido_etal92,Mukherjee&Kaplan95}.
In particular, the lagged cell's response shows a phase lag relative to the
input that increases with frequency; the nonlagged cell goes through a
transition between a low-pass and in-phase relay mode to a band-pass and
phase-lead (at frequencies $< 8$ Hz) relay mode as the membrane hyperpolarizes.
The former corresponds to the tonic, the latter to the burst relay mode.

\begin{figure}
\begin{minipage}{0.5\textwidth}
\leftline{\includegraphics[width=0.9\textwidth]{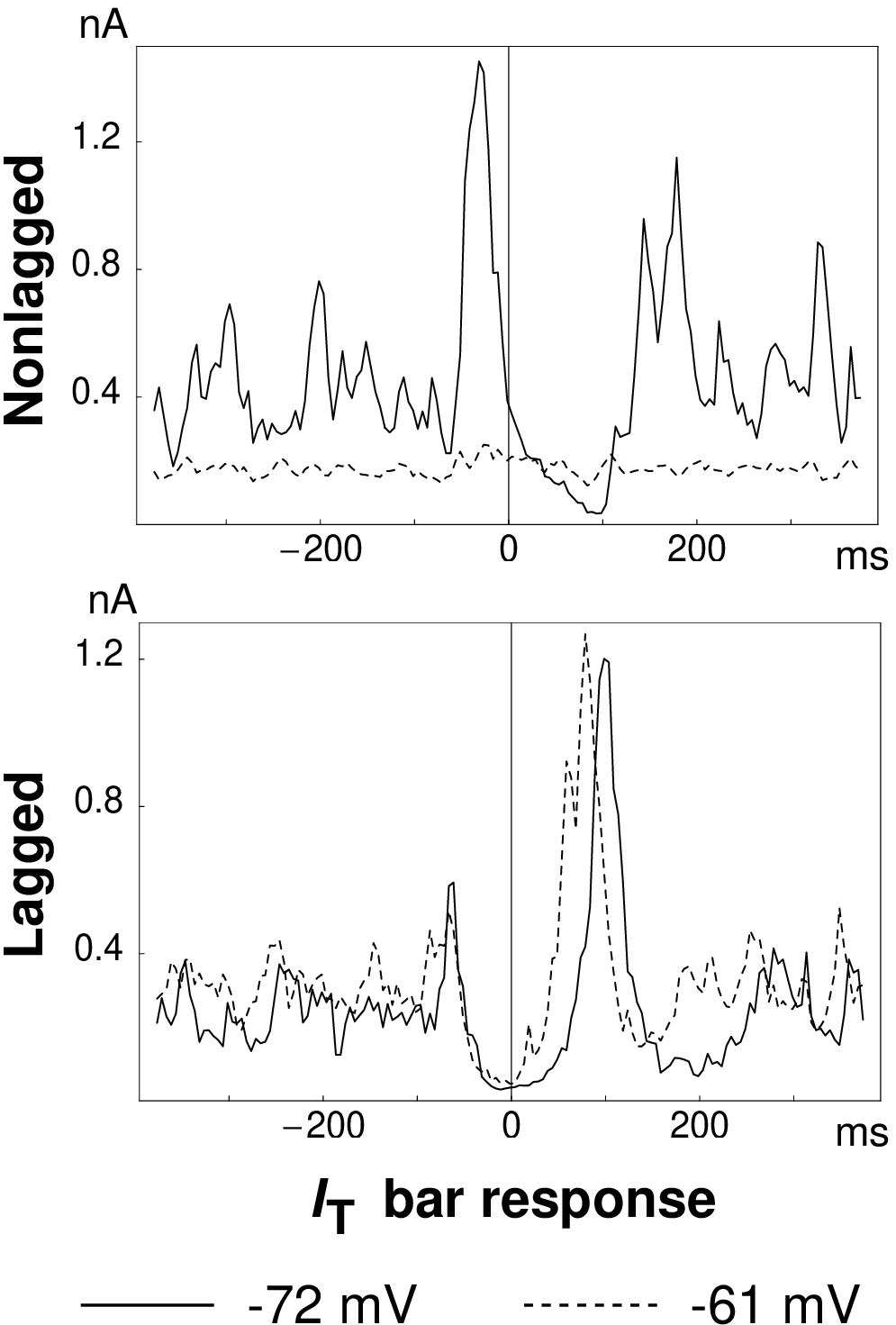}}
\end{minipage}
\begin{minipage}{0.49\textwidth}
\caption[]{Transient and low-threshold Ca$^{2+}$ current $I_{\rm T}$ associated
with the bar-stimulus scenarios shown in Figure \ref{61_72_mV} leftmost column
(averaged over 100 bar sweeps). High $I_{\rm T}$ indicates burst spikes
mediated by underlying Ca$^{2+}$ spikes. For the nonlagged neuron at a
resting membrane potential of $-$61 mV, $I_{\rm T}$ is always small and does not
contribute to the response. In the remaining cases, the timing of the response
shown in the leftmost column of Figure \ref{61_72_mV} can be seen to be largely
determined by $I_{\rm T}$.}
\label{IT_61_72_mV}
\end{minipage}
\end{figure}

Remarkably, as the resting membrane potential is varied, the timing of the bar
response shifts in {\em opposite} directions for lagged and nonlagged cells;
cf.\ Figure \ref{61_72_mV} left column. Increasing hyperpolarization shifts the
lagged response peak to {\em later} times, while the nonlagged response peak
moves to {\em earlier} times. In view of what we have reported in section
\ref{I.2} on relay modes and lagged cells, it seems likely that the
low-threshold Ca$^{2+}$ current $I_{\rm T}$ is in part responsible for the
GRCs' response timing. In Figure \ref{IT_61_72_mV} we show simulated traces of
$I_{\rm T}$ for the moving-bar scenarios. For nonlagged neurons, the current is
insignificant at $-$61 mV, but exhibits a pronounced peak at the start of the
response to the bar at $-$72 mV. The peak of $I_{\rm T}$ confirms the nature of
the early response component seen in Figure \ref{61_72_mV} top left column as
Ca$^{2+}$-mediated burst spikes, in agreement with Lu et al.\ (1992), Guido et
al.\ (1992), and Mukherjee \& Kaplan
(1995)\nocite{Lu_etal92,Guido_etal92,Mukherjee&Kaplan95}. For lagged neurons,
on the other hand, we see that the timing of the $I_{\rm T}$ peak faithfully
reflects the timing of the response peak at both resting membrane potentials;
see Figure \ref{61_72_mV} bottom left column. In fact, the profile of the
$I_{\rm T}$ traces resembles the one of the spike rates, indicating that the
Ca$^{2+}$ current promotes firing throughout the transient responses simulated
here. We will return to the significance of burst spikes for the results in
section \ref{Discussion}.

The reason for the opposite shifts of lagged and nonlagged response timing,
then, lies in the interaction of the low-threshold Ca$^{2+}$ current $I_{\rm
T}$ with the different levels of inhibition received by lagged and nonlagged
neurons. With only weak feedforward inhibition, nonlagged neurons respond to
retinal input with immediate depolarization, eventually reaching the activation
threshold for the Ca$^{2+}$ current. If the Ca$^{2+}$ current is in the
de-inactivated state, it will boost depolarization and give rise to an early
burst component of the visual response. The lower the resting membrane
potential, the more de-inactivated and, hence, stronger the Ca$^{2+}$ current
will be, and the stronger the early burst relative to the late tonic response
component. Lagged neurons, on the other hand, receive strong feedforward
inhibition and, hence, initially respond to retinal input with
hyperpolarization. Repolarization occurs when inhibition gets weaker. This may
result either from cessation of retinal input or from adaptation, i.e.,
fatigue, of the inhibitory input to GRCs; cf.\ Figure \ref{model_graphs}{\bf
A}. With the Ca$^{2+}$ current $I_{\rm T}$ being de-inactivated by the
excursion of the membrane potential to low values, lagged spiking starts with
burst spikes as soon as the voltage reaches the Ca$^{2+}$-activation threshold.
This will take longer, if the resting membrane potential is lower, leading to
the shift in response timing with membrane polarization observed here.

Adaptation of inhibition is implemented in the present model by the
refractoriness of the spike-response neurons that represent the inhibitory
interneurons; cf.\ section \ref{Gen_Input}. The refractory potential saturates,
however, on a timescale (10.0 ms) much shorter than the delay of the lagged
response of roughly 100 ms; cf.\ Figure \ref{61_72_mV}. Its role in generating
a response delay for lagged neurons in our model can thus be only very limited.

It is important to note that the lagged on-response is different from a
nonlagged off-response. A nonlagged off-response produces a phase lag of half a
cycle relative to the nonlagged on-response at all frequencies. The right
column of Figure \ref{61_72_mV} shows that this is not true for the simulated
lagged response. Rather, the phase-transfer function has a significantly higher
slope -- that is, a higher phase latency -- for the lagged response than for
the nonlagged response (cf.\ Saul \& Humphrey, 1990)\nocite{Saul&Humphrey90} at
both resting membrane potentials. Moreover, we observed that lagged cells
produce a delay of moving-bar responses that does not vanish at high speeds
(not shown). This fixed delay component must be largely determined by the
internal neuronal dynamics of the ion currents, notably of $I_{\rm T}$, that
follows hyperpolarization.

For the remaining simulations we have always set the peak postsynaptic
conductances for lagged and nonlagged neurons to the values used for the data
shown in Figures \ref{61_72_mV} and \ref{IT_61_72_mV}.

\subsection{Total Geniculate Input to Cortex}

\begin{figure}
\centerline{\includegraphics[width=\textwidth]{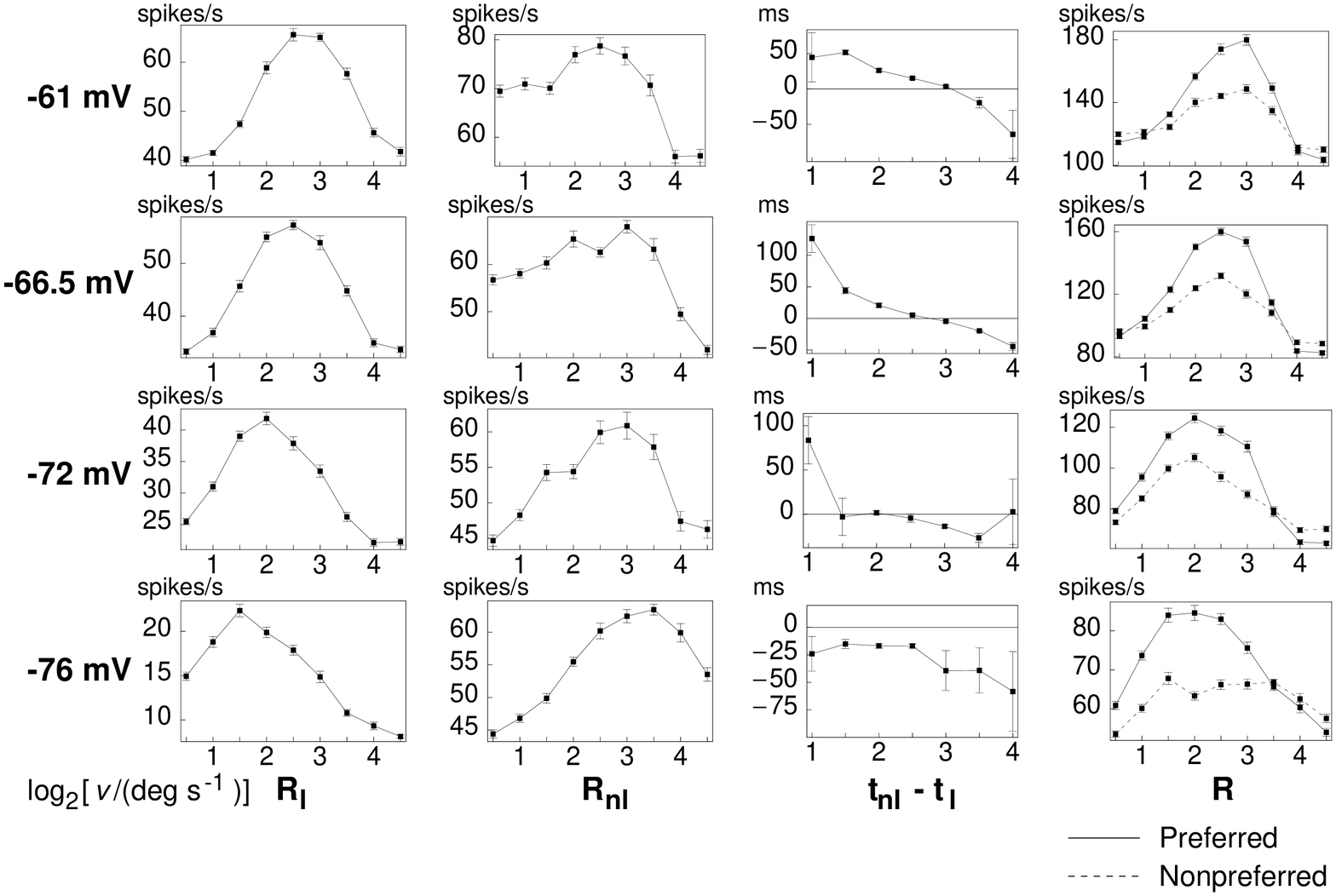}}
\caption[]{Geniculate moving-bar response and geniculate input to cortex as
predicted by model simulations. Velocity tuning and timing of the response
peaks have been plotted for resting membrane potentials indicated on the far
left. In the columns we show from left to right as functions of the bar
velocity the peak response rate of the lagged population (${\rm R}_{\rm l}$),
of the nonlagged population (${\rm R}_{\rm nl}$), their peak-time difference
(${\rm t}_{\rm nl} - {\rm t}_{\rm l}$) for the preferred direction, and the
total geniculate input (R) to a cortical cell for the preferred (solid lines)
and nonpreferred (dashed lines) direction of motion. The horizontal axes show
the logarithm (base 2) of speed in all graphs. The bars in the graphs are
standard errors. As the membrane is {\em hyperpolarized}, the total geniculate
input to a cortical cell peaks at progressively {\em lower} velocities. Means
and standard errors are estimated from 30 bar sweeps. Figure adapted from
Hillenbrand \& van Hemmen (2000)\nocite{Hillenbrand&vanHemmen00}.}
\label{LGN_dats}
\end{figure}

Lagged and nonlagged responses have to be combined so as to yield a
velocity-selective input to a cortical neuron. For different values of the
resting membrane potential, Figure \ref{LGN_dats} shows in the columns from
left to right the velocity tuning of the lagged population (${\rm R}_{\rm l}$),
of the nonlagged population (${\rm R}_{\rm nl}$), the peak-time differences
(${\rm t}_{\rm nl} - {\rm t}_{\rm l}$) of their responses for the preferred
direction, and the tuning of the total geniculate input (R) to a cortical cell
for the preferred and nonpreferred direction of motion; see section
\ref{DataAnalysis} for details. As in vivo, the lagged cells prefer lower
velocities and have lower peak firing rates than the nonlagged cells
\cite{Mastronarde87a,Humphrey&Weller88a,Saul&Humphrey90}. The key observation,
however, is that the maximum of the {\em total geniculate input rate} to a
cortical neuron shifts to {\em lower velocities} as the membrane potential {\em
hyperpolarizes}; see Figure \ref{LGN_dats} right column.

The total geniculate input rate R assumes its maximum at a velocity of bar
motion where the peak discharges of the lagged and nonlagged neurons {\em
coincide}, i.e., where ${\rm t}_{\rm nl} - {\rm t}_{\rm l} \approx 0$. The
shift of the maximum with hyperpolarization to lower velocities is produced by
a corresponding shift of the peak-time differences ${\rm t}_{\rm nl} - {\rm
t}_{\rm l}$ and of the lagged tuning ${\rm R}_{\rm l}$, while the maximum of the
nonlagged tuning ${\rm R}_{\rm nl}$ remains essentially unchanged. The shift of
the peak-time differences, in turn, is a reflection of the opposite shifts in
bar-response timing of lagged and nonlagged neurons described in the previous
subsection; cf.\ Figure \ref{61_72_mV} left column. The total geniculate input
rate R is higher for the direction of bar motion where ${\rm t}_{\rm nl} - {\rm
t}_{\rm l}$ assumes lower values. In other words, the direction preferred is
the one where the lagged cells receive their retinal input before the nonlagged
cells; cf.\ Figures \ref{model_graphs}{\bf B} and \ref{LGN_dats} right column.

We found that feedback inhibition from the PGN does not affect the timing of
lagged and nonlagged responses. Its only effect is to reduce variations in
response amplitude by countering increases in firing rate of relay neurons with
stronger inhibition. The PGN feedback loop thus moderates the differences in
response activity both at different levels of the resting membrane potential
and between lagged and nonlagged neurons. The latter difference may be further
reduced by making the feedback inhibition stronger for nonlagged than for
lagged neurons. For the data shown in Figure \ref{LGN_dats} this has been
implemented by allowing stronger or, equivalently, more synapses of nonlagged
neurons on PGN cells than synapses of lagged neurons. Despite this, the
nonlagged responses dominate and there is a drop of geniculate activity,
especially of the lagged responses, with increasing hyperpolarization. We will
return to this issue in section \ref{Discussion}. In general, however, the PGN
loop increases the range of resting membrane potentials of relay cells that
yield balanced lagged and nonlagged inputs to the cortex, thereby extending the
dynamic range of speed tuning of the total geniculate input to a cortical
neuron.

\section{Discussion}
\label{Discussion}

Recently it has been proposed that corticogeniculate feedback modulates the
spatial layout of simple-cell RFs by exploiting the thalamic burst-tonic
transition of relay modes \cite{Worgotter_etal98}. Along a similar line, the
main point made by our modeling is that one should expect a modulatory
influence of cortical feedback on the spatio{\em temporal} RF structure of
simple cells. More precisely, we observe a shift in the time to the
bar-response peak that is {\em opposite} for lagged and nonlagged cells; cf.\
Figure \ref{61_72_mV} left column. Assuming (i) an RF layout as usually found
for direction-selective simple cells in area 17, and (ii) an influence of {\em
convergent} geniculate lagged and nonlagged inputs on this RF structure, it
follows that the observed shifts in response timing affect cortical speed
tuning. To the best of our knowledge, nobody has looked for such an effect yet.

We have investigated  the geniculate input to simple cells, which clearly
cannot be compared with their output directly. Because of intracortical
processing we cannot expect to reproduce tuning widths and direction
selectivity indices of cortical neurons. Rather, the tuning width of geniculate
input is likely to be larger, and its directional selectivity weaker than of a
cortical neuron's output; cf.\ section \ref{I.1}. Indeed, superficial
inspection of the rightmost column of Figure \ref{LGN_dats} reveals that the
directional bias of R is rather weak compared to what can be found for
directional cells in cat areas 17 and 18 \cite{Orban_etal81b}. On the other
hand, the tuning width of R is relatively narrow \cite{Orban_etal81a}, instead
of wide. The narrowness of speed tuning in our simulations may be reconciled
with experimental data in the following ways. First, we have simulated the
ideal case of equal resting membrane potential, and hence lagged and nonlagged
response timing, for all of the GRCs. Scattered values of membrane potentials
will produce less sharply tuned profiles for R. Second, if it is true that
velocity tuning is not a {\em static} but a {\em dynamic} property of cortical
cells, as is proposed in this article, measured -- {\em effective} -- tuning
widths should be larger than the width of the tuning under static conditions as
simulated here.

Quantitative comparison of the tuning of R with cortical velocity tuning is,
for the above reasons, problematic. Nonetheless it is interesting to note that,
much like velocity tuning in areas 17 and 18 \cite{Orban_etal81a}, the dynamic
range of the modeled geniculate input -- that is, the difference between the
highest and the lowest response values on each tuning curve ${\rm R}(v)$ --
decreases and the tuning width increases with decreasing optimal
velocity\footnote{The correlation with tuning width was significant only in
area 18 \cite{Orban_etal81a}.}; see Figure \ref{LGN_dats} right column.
Moreover, the range of preferred velocities lies within the range observed for
velocity-tuned cells \cite{Orban_etal81a}.

Because of scaling properties of the retinal ganglion cells' velocity tuning
\cite{Cleland&Harding83}, rescaled versions of the RF geometry shown in Figure
\ref{model_graphs}{\bf B} produce accordingly shifted tuning curves (on a
logarithmic speed scale). In particular, we retrieve the positive correlation
between RF size and preferred speed found in areas 17 and 18
\cite{Orban_etal81a} from the geniculate input.

The effects of lagged and nonlagged response timing in the present model are
dependent on the low-threshold Ca$^{2+}$ current and ensuing burst spikes.
The significance of our results for visual processing in the awake, behaving
animal, then, is subject to the occurrence of burst spikes under such
conditions. As mentioned in section \ref{I.2}, this issue is still under much
debate. For nonlagged cells, burst spikes will have a role in normal vision
only, if their resting membrane potential gets hyperpolarized enough. For
lagged cells, it is presently not settled, if their (transient) responses are
indeed supported by the low-threshold Ca$^{2+}$ current, as was seen in the
simulations. If this turns out to be wrong, the effect of cortical input on
lagged response timing could be different from what we have observed. In this
regard, it would be interesting to study the effect of additional NMDA channels
at the synapses of retinal afferents on GRCs; cf.\ Heggelund \& Hartveit (1990)
and Hartveit \& Heggelund
(1990)\nocite{Heggelund&Hartveit90,Hartveit&Heggelund90}. Nonetheless, the data
on response timing of the modeled lagged cells suggest that some essential
aspect of the true lagged mechanism has been captured in the model.

Responses of X-relay cells to moving bars and textures are on average reduced
after ablation of the visual cortex in cats \cite{Gulyas_etal90}. This is
consistent with what we observe in our simulations of relay cells, assuming a
depolarizing net effect of cortical feedback on relay neurons
\cite{Funke&Eysel92,Worgotter_etal98}. In fact, the response rates of lagged
and nonlagged neurons decrease with progressive hyperpolarization (cf.\ Figure
\ref{LGN_dats} first and second columns), despite disinhibition by PGN
feedback.

The question arises of how the visual cortex would deal with the resulting
differences in the maximal geniculate input activity (cf.\ Figure
\ref{LGN_dats} right column) in a way that preserves the speed tuning of the
afferent signal for a wide range of geniculate membrane polarizations. In
principle this is straightforward since it is area 17 itself that modulates the
membrane potential of relay cells. By a similar mechanism it could likewise
adjust the responsiveness of layer 4B neurons to geniculate input. An
appropriate modulatory signal could most easily be derived from the same layer
6 neurons that project to the LGN, or from their neighbors that share the same
information on the actual corticothalamic feedback. In this context it is very
interesting that layer 6 neurons that project to the LGN indeed send axon
collaterals specifically to layer 4 \cite{Katz87}.

The PGN, and more generally the thalamic reticular nucleus, implements both a
disynaptic inhibitory feedback loop and an indirect corticothalamic feedback
pathway to relay cells \cite{Sherman96,Sherman&Guillery96}. This double role
suggests possible interactions between the two functions. Depending on whether
or not individual PGN neurons engage in both types of circuitry and on the
details of connections between different PGN neurons, the strength of the
disynaptic feedback inhibition exerted by PGN neurons on GRCs could be
modulated by cortical feedback. Unlike in our simulations, the efficiency of
the LGN-PGN loop might thus {\em covary} with the GRCs' resting membrane
potential. In theory this would offer a very elegant mechanism to compensate
the above-mentioned differences in GRC-response level at different resting
membrane potentials. For the time being this is, of course, mere speculation.

Recently, it has been found that responses in the thalamic reticular nucleus of
rat that are mediated by a specific subtype of metabotropic glutamate receptor
(group II) result in long-lasting cell hyperpolarization \cite{Cox&Sherman99},
instead of depolarization as usually. The effect seems to be caused by opening
of a K$^+$ leak channel similar to a GABA$_{\rm B}$ response. This observation
adds some more variants of possible corticothalamic pathways for the slow
control of thalamic membrane potential. Specifically, it suggests that relay
cells may be depolarized by reticular disinhibition. Moreover, if group II
receptors turned out to be active on relay neurons as they are on reticular
neurons, a direct hyperpolarizing effect of corticothalamic feedback would
become conceivable.

In the model we have considered only one type of cortical input to the LGN,
namely, the input mediated by metabotropic receptors that slowly control a
K$^+$ leak conductance on GRCs; cf.\ section \ref{CTX_feedback}. There are other
cortical inputs, mediated by ionotropic receptors, that act on the much shorter
timescale of the retinal inputs. While such cortical feedback certainly
influences the detailed temporal pattern of geniculate spiking [see, e.g.,
Sillito et al.\ (1994)\nocite{Sillito_etal94}], it seems unlikely that they
affect the gross timing of a transient response peak on a timescale of several
10 ms. An interesting exception is perhaps NMDA receptor-mediated feedback,
with time constants in-between those of metabotropic and (ionotropic)
AMPA/kainate or GABA$_{\rm A}$ responses. In future work it would be,
therefore, interesting to include NMDA channels at corticothalamic synapses in
the model.

We have presented arguments for the existence of a particular dynamic gating
mechanism for thalamocortical information transfer, namely, for the transfer of
information on visual motion. New experiments are required to check the
implications directly. If the proposed mechanism turns out to be effective in
awake, behaving animals, it will have important, as yet unrecognized,
consequences for motion processing. A possible implication in motion-mediated
object segmentation is discussed in Hillenbrand \& van Hemmen
(2000)\nocite{Hillenbrand&vanHemmen00}.

\renewcommand{\theequation}{\Alph{section}.\arabic{equation}}
\setcounter{section}{1}
\setcounter{equation}{0}

\section*{Appendix}

We here give a brief account of essential concepts that are related to
biophysical neuron models and, in particular, to the model of the thalamic
relay neuron \cite{Huguenard&McCormick92,McCormick&Huguenard92} studied in this
work. For a detailed exposition of data and theory on ion channels and
excitable membranes the reader is referred to Tuckwell (1988a,
1988b)\nocite{Tuckwell88a,Tuckwell88b} and Hille (1992)\nocite{Hille92}.

The essential electrical properties of neuronal membranes are described by the
differential equation
\beq
\label{dV/dt}
\frac{{\rm d}V}{{\rm d}t} \, = \, \frac{1}{C} \sum_{i=1}^n I_i \ ,
\eeq
where $V$ is the cell's membrane potential, $I_i$ are the currents through the
different types of ion channels in the membrane, and $C$ is the membrane
capacitance. The art of building a neuron model is to find good empirical,
quantitative descriptions of all the relevant ion currents. Equation
\ref{dV/dt} describes a point-like neuron or a single compartment of an
extended neuron. Thalamic relay neurons are well described by
single-compartment models \cite{Huguenard&McCormick92,McCormick&Huguenard92}.

Within the Ohmic approximation, the ion currents are described by
\beq
\label{IionOhmic}
I_i \, = \, g_i \, m_i^{p_i} \, h_i^{q_i} \, \left( V_i - V \right) \ ,
\eeq
with the {\em reversal potential} $V_i$, the maximal conductance $g_i$, the
{\em gates} $m_i$ and $h_i$, and some positive, usually integer, constants
$p_i$ and $q_i$. The reversal potential $V_i$ is approximately equal to the
Nernst potential for ions of type $i$ but is usually determined empirically.
The gates $m_i$ and $h_i$ are dynamic variables that assume values between zero
and one according to differential equations that involve the membrane potential
$V$; see below.

It must be stressed that expressions of type \ref{IionOhmic} are primarily
{\em empirical fits} to the voltage dependence of ionic currents. Nonetheless,
an oversimplified but intuitive physical interpretation of \ref{IionOhmic} is
that ion currents flow through an ensemble of channels of type $i$ that have
$p_i$ $m$-gates and $q_i$ $h$-gates each. The gates are open and closed with
certain probabilities. An individual channel allows ions to pass only if {\em
all} its gates are in the open state.

In what follows we will drop the index $i$ for notational simplicity. With the
given picture of ionic gates in mind, we may `understand' the dynamics of the
gates $m$ and $h$. Transitions between the open and closed states are governed
by the transition rates $\alpha_{m/h}$ and $\beta_{m/h}$,
\bea
\label{dm/dt}
\frac{{\rm d}m}{{\rm d}t} & = & \alpha_m(V) \, (1 - m) - \beta_m(V) \, m \ ,\\
\label{dh/dt}
\frac{{\rm d}h}{{\rm d}t} & = & \alpha_h(V) \, (1 - h) - \beta_h(V) \, h \ .
\eea
The rates, in turn, are functions of the membrane potential $V$. Instead of
transition rates, one may specify the gates' asymptotic values $m_{\infty}$ and
$h_{\infty}$, and time constants $\tau_{m/h}$. Their relation to the transition
rates is
\bea
m_{\infty}(V) & = & \frac{\alpha_m(V)}{\alpha_m(V) + \beta_m(V)} \ ,\\
h_{\infty}(V) & = & \frac{\alpha_h(V)}{\alpha_h(V) + \beta_h(V)} \ ,\\
\tau_{m/h}(V) & = & \frac{1}{\alpha_{m/h}(V) + \beta_{m/h}(V)} \ .
\eea
By convention, the $m$-gate is usually the one that opens ($m_{\infty} \approx
1$) at higher and closes ($m_ {\infty}\approx 0$) at lower membrane potentials;
for the $h$-gate the situation is just the other way round. The $m$-gate is
called the {\em activation gate}, the $h$-gate the {\em inactivation gate}.
Accordingly, a current is said to {\em activate} when the $m$-gate opens; it is
said to {\em inactivate} when the $h$-gate closes.

For the firing pattern of thalamic relay neurons, the transient and
low-threshold Ca$^{2+}$ current $I_{\rm T}$ is of particular importance.
Analogous to the production of Na$^+$ spikes by the transient Na$^+$ current
$I_{\rm Na}$, $I_{\rm T}$ produces Ca$^{2+}$ spikes that, in turn, can promote
Na$^+$ spikes; see section \ref{I.2}.

Some types of ion channels do not inactivate, i.e., they have $q = 0$. An ion
channel that neither inactivates nor de-activates, i.e., $p = q = 0$, is called
a {\em leak} channel. Leak channels are characterized by a constant conductance
$g$; cf.\ equation \ref{IionOhmic}.

For some ion channels the Ohmic approximation \ref{IionOhmic} for the ion
current $I$ is not satisfactory. In those cases Goldman's constant-field
equation
\beq
\label{IionConstantField}
I \, = \, g \, m^p \, h^q \, \frac{V z^2 e^2}{k T} \, \frac{c_{\rm i} - c_{\rm e} \exp(-z e V/k T)}{1 - \exp(-z e V/k T)}
\eeq
often is a better choice \cite{Tuckwell88a}. Here $c_{\rm i}$ and $c_{\rm e}$
are the ion's concentrations in the intra- and extracellular space,
respectively, and $z$ is its valence. As usually, $e$ is the elementary charge,
$k$ is the Boltzmann constant, and $T$ is the absolute temperature. For the
thalamic relay neuron, the Ca$^{2+}$ currents are modeled according to equation
\ref{IionConstantField}.

Besides the voltage-gated channels introduced here, thalamic relay neurons have
channels that are gated by membrane potential {\em and} the intracellular
concentration of Ca$^{2+}$ ions
\cite{Huguenard&McCormick92,McCormick&Huguenard92}. Their transition rates
$\alpha$ and $\beta$ [cf.\ equations \ref{dm/dt} and \ref{dh/dt}] are functions
of membrane voltage and intracellular Ca$^{2+}$ concentration. Moreover,
receptor-gated channels are responsible for most of the synaptic transmission
in the central nervous system; cf.\ equation \ref{g(t)}.

\section*{Acknowledgments}

We thank Esther Peterhans and Christof Koch for stimulating discussions. U.H.\
was supported by the Deutsche Forschungsgemeinschaft (DFG), grant GRK 267.

\end{document}